\documentclass[lettersize,journal]{IEEEtran}

\usepackage[T1]{fontenc}
\usepackage[dvipsnames,svgnames,x11names]{xcolor}

\usepackage[utf8]{inputenc}
\usepackage{acronym}
\usepackage{graphicx}
\usepackage{float}
\usepackage{tikz}
\usepackage{framed}
\usepackage{gensymb}
\usepackage[colorinlistoftodos,prependcaption,textsize=footnotesize]{todonotes}

\usepackage{multirow}
\usepackage[hyphens]{url}
\usepackage{hyperref} 

\usepackage{graphicx} 

\usepackage[english]{babel}

\usepackage[markup=underlined]{changes}
\definechangesauthor[color=BrickRed]{SS}
\definechangesauthor[color=NavyBlue]{FM}
\usepackage{comment}

\acrodef{ADC}{Analog-to-Digital Converter}
\acrodef{LTE}{Long Term Evolution}
\acrodef{KPI}{Key Performance Indicator}
\acrodef{PLC}{Power Line Communication}
\acrodef{IoT}{Internet of Things}
\acrodef{SG}{Smart Grid}
\acrodef{HIL}{Hardware-in-the-Loop}
\acrodef{SE}{state estimation}
\acrodef{MAE}{Mean Absolute Error}
\acrodef{RMSE}{Root Mean Square Error}
\acrodef{RPi}{Raspberry Pi}
\acrodef{PDC}{Phasor Data Concentrator}
\acrodef{NTP}{Network Time Protocol}
\acrodef{PMU}{Phasor Measurement Unit}
\acrodef{GPS}{Global Positioning System}
\acrodef{mMTC}{machine-type communication}
\acrodef{WLS}{weighted least squares}
\acrodef{MAE}{Mean Absolute Error}
\acrodef{RMSE}{Root Mean Square Error}
\acrodef{URLLC}{ultra-reliable low-latency communications}
\acrodef{eMBB}{enhanced mobile broadband}
\acrodef{WAMS}{Wide Area Monitoring Systems}
\acrodef{COTS}{commercial off-the-shelf}
\acrodef{BS}{Base Station}
\acrodef{fps} {frames per second}

\usepackage{cite}

\ifCLASSINFOpdf
\else
\fi
\usepackage{amsmath}
\usepackage[cmintegrals]{newtxmath}
\usepackage{bm}

\usepackage{datetime}
\usepackage{caption}

\ifCLASSOPTIONcompsoc
  \usepackage[caption=false,font=normalsize,labelfont=sf,textfont=sf,subrefformat=parens,labelformat=parens]{subfig}
\else
  \usepackage[caption=true,font=footnotesize,subrefformat=parens,labelformat=parens]{subfig}
\fi
\usepackage{url}

\usepackage{authblk}

\title{Real-Time State Estimation in Smart Grids over 5G Networks: Experimental Validation Using Raspberry Pis and Typhoon HIL}
 \author[]{Biswajit~Kumar~Dash}
 \author[]{Luis~Herrera}
 \author[]{Filippo~Malandra}
\affil[]{Department of Electrical Engineering, State University of New York at Buffalo, NY, USA\\
Email: \{biswajit, lcherrer, filippom\}@buffalo.edu}

\IEEEoverridecommandlockouts
\begin{document}

\maketitle

\begin{abstract}
Reliable, low-latency communication is critical for real-time monitoring and control in modern Smart Grids (SGs). The emergence of 5G networks, with enhanced reliability, significantly lower latency, and native support for massive machine-type communication, offers strong potential to enable advanced grid applications such as state estimation (SE) and fault detection. While existing studies investigate 5G for SG use cases, most rely on simulations or analytical models; experimental validation using real hardware and SG data remains limited. This paper fills this gap by presenting a fully experimental validation of real-time SE over a commercial 5G network using a 5G-based multi-node testbed built with Raspberry Pi (RPi)-based SG nodes and a Typhoon Hardware-in-the-Loop (HIL) real-time simulator. We first characterize 5G communication performance using simulated SG data under varying reporting rates and deployment environments by evaluating Key Performance Indicators (KPIs) such as end-to-end delay, jitter, and frame loss. Experimental results show that the worst-case mean delay observed for the 5G is approximately 6.5$\times$ lower than that of our previous LTE cat-M study at the corresponding reporting rate. We then stream real-time voltage, current, and phase-angle measurements---generated by an IEEE 4-node feeder model in Typhoon HIL simulator---to a remote Phasor Data Concentrator (PDC) for SE and fault detection. Results demonstrate that 5G-enabled measurements support accurate SE under both steady-state and dynamic load variations. Furthermore, fault-detection experiments confirm reliable and prompt fault detection, with detection delays as low as 0.80\,s. 
\end{abstract}

\begin{IEEEkeywords}
Smart Grid communication, 5G Networks, Multi-node testbed, Hardware-in-the-Loop (HIL), Network performance, State Estimation, Synchrophasor Measurements
\end{IEEEkeywords}
\section{Introduction}
\IEEEPARstart{E}{merging} and future \ac{SG} systems increasingly rely on advanced communication and networking technologies to meet the growing complexity and diversity of modern power systems. With global shifts towards sustainable energy sources and the adoption of diverse \acs{SG} applications, there is a pressing need for enhanced reliability, reduced latency, and scalable connectivity. These improvements are crucial for enabling efficient, bidirectional data exchange across the distributed power grid, thereby supporting intelligent monitoring, estimation, and decision-making capabilities.

The growing demand for reliable, high-performance, and scalable communication solutions in \acs{SG} applications has driven the adoption of 5G technology, built upon its three foundational pillars: \ac{eMBB}, \ac{URLLC}, and \ac{mMTC}~\cite{cosovic20175g}. These capabilities promise to meet the stringent requirements of SG functions such as real-time monitoring, distributed control, and \ac{SE}, which are critical for ensuring grid stability and resilience. Despite these promises, the practical deployment of 5G in \ac{SG} environments requires rigorous performance analysis to validate its ability to deliver low latency, high reliability, and scalability under realistic conditions. Existing studies have primarily relied on three approaches: mathematical analysis (e.g.,~\cite{Zerihun2020}), which offers theoretical insights but often oversimplifies network behavior; network simulation~\cite{cosovic20175g, Cosovic2018}, which enables large-scale scenario modeling but lacks hardware-level fidelity; and field trials (e.g.,~\cite{holfeld2015smart}), which provide real-world validation but are costly and difficult to scale. Moreover, these methods rarely capture the tight coupling between communication performance and power system dynamics, leaving a critical gap in understanding the end-to-end behavior of SG applications over 5G networks.

To address this gap, experimental performance analysis has emerged as a promising alternative~\cite{10143367, 10333918, 11204602}. Unlike purely simulated environments, experimental setups allow direct measurement of latency, throughput, and reliability while accounting for hardware constraints, protocol overheads, and environmental factors. More importantly, by integrating communication hardware with real-time power system simulators such as Typhoon \ac{HIL} or OPAL-RT, researchers can evaluate how network performance affects essential grid functions, such as \acs{SE} and control~\cite{kharchouf2024controller}. This holistic approach provides actionable insights for system operators and technology developers. However, experimental evaluation introduces its own challenges, including the complexity of integrating heterogeneous hardware, ensuring synchronization between communication and power domains, and maintaining reproducibility across different test conditions.

In this work, we present a multi-node experimental testbed that leverages \ac{COTS} devices, such as \ac{RPi} units equipped with 5G connectivity, integrated with a Typhoon HIL real-time simulator. This architecture enables the execution of \acs{SE} algorithms over a commercial U.S.-based 5G communication network while maintaining tight coupling with power system dynamics. By combining cost-effective hardware with high-fidelity simulation, our approach delivers a scalable and reproducible platform for evaluating 5G performance in SG applications. The results demonstrate the feasibility of deploying 5G for critical SG functions and provide valuable insights into latency, reliability, and integration challenges.

The main contributions of this work are as follows:
\begin{itemize}
    \item Design and deployment of a multi-node testbed that integrates 5G connectivity with high-fidelity real-time power system simulation with Typhoon \acs{HIL}.
    \item Experimental evaluation of 5G network performance to study its feasibility for a variety of \ac{SG} applications.
    \item Validation of real-time \acs{SE} over a commercial 5G network using synchrophasor data from an IEEE 4-node feeder.
    \item Demonstration of fault-detection capability by injecting faults into the test feeder and showing that a global residual-based metric can reliably and quickly identify fault events.
\end{itemize}

The remainder of this paper is organized as follows. Section~\ref{section:SOA} reviews the state of the art. Section~\ref{section:system} describes the architecture of our 5G-enabled \acs{SG} testbed. Section~\ref{sec:DelayAnalysis} presents experimental results on the 5G communication performance. Section~\ref{section:SE} provides the real-time \acs{SE} framework and experimental validation over 5G networks. Finally, Section~\ref{section:Conclusion_SE} concludes the paper.

\section{State of the Art}\label{section:SOA}
This section reviews the existing literature on communication technologies for \acp{SG} in Section~\ref{subsec:Commfor_SG}, discusses how communication performance influences key \acs{SG} applications in Section~\ref{subsec:ImpactofComm_SG}, and identifies the research gaps that motivate this work.

\subsection{Communication Technologies for Smart Grids} \label{subsec:Commfor_SG}

Communication networks play a critical role in supporting monitoring, control, and automation in modern \acp{SG}~\cite{srivastava22, suhaimy22}. Although both wired and wireless technologies have been explored to meet the stringent latency, reliability, and scalability demands of emerging grid applications, wireless networks become attractive due to their flexibility, scalability, and lower installation cost~\cite{appasani2018review}.


A variety of wireless technologies, including WLANs and Zigbee, has been examined for distribution-level monitoring and control~\cite{gharavi2015scalable,5975716}. Although these systems demonstrate useful capabilities, their reliability, coexistence performance, and coverage limitations restrict large-scale deployment in \acs{SG} environments. Zigbee, for instance, is particularly vulnerable in harsh or interference-heavy environments~\cite{dash2022zigbee}. Broadband cellular networks such as 3G and LTE have also been evaluated for grid telemetry~\cite{8580965}, with studies considering both machine-to-machine (M2M) and mixed traffic scenarios~\cite{malandra2017case, malandra2018traffic}. However, LTE is not optimized for transmitting small, periodic sensor measurements due to its heavy protocol overhead~\cite{samoilenko20}.

Studies have therefore explored cellular IoT technologies such as LTE cat-M and NB-IoT, which offer reduced device costs, improved coverage, and lower energy consumption~\cite{8422888,7983347}. Our previous work experimentally evaluated LTE cat-M for \acs{SG} communications using a single Arduino-based measurement node~\cite{10143367} and later extended the study to a multi-node configuration~\cite{10333918}. These studies highlighted two practical limitations. First, the measured delay (approximately 176--200\,ms) exceeded the latency requirements of several key applications, including \ac{SE}. Second, these studies primarily evaluated communication performance using \textit{simulated} measurement data, leaving open questions about system performance with actual grid signals. 

These limitations motivate the exploration of emerging 5G networks, which promise significantly lower latency, higher reliability, and improved support for massive deployments of lightweight \acs{SG} devices.

\subsection{Impact of Communications on Smart Grid Applications} \label{subsec:ImpactofComm_SG}

The high reliability and low latency of 5G communications make them ideal for time-critical \acs{SG} applications, including \acs{SE}, fault detection, and control that require timely and accurate measurement data. In~\cite{porcu20215g}, the authors examine the use of 5G technologies for fault identification, estimation, monitoring, and fault distributed generation control. The impact of 5G communication failures on \ac{SE} techniques is investigated in~\cite{Zerihun2020}. Leveraging the low-latency and high-reliability properties of 5G, \cite{cosovic20175g} proposes distributed \ac{SE} methods using the Alternating Direction Method of Multipliers (ADMM) and Belief Propagation (BP). Similarly, \cite{Cosovic2018} develops a Gaussian Belief Propagation (GBP) method for linear \acs{SE} in 5G cloud radio access networks. 

Beyond 5G-focused studies, several works analyze how communication disruptions affect \ac{SE} regardless of the underlying technology. Švenda et al.~\cite{vsvenda2018influence} analyze the impact of delays and packet loss on \acs{SE} accuracy and show that such irregularities can significantly increase the mean square error in voltage angle estimations. Cokic et al.~\cite{cokic2021communication} highlight how network congestion degrades real-time application performance by increasing latency and delivering outdated information. Tsitsimelis et al.~\cite{tsitsimelis2018impact} examine the reliability of LTE's random-access channel (RACH) and demonstrate how increased device contention and cell coverage variations can impair \ac{WAMS}. Aminifar et al.\cite{aminifar2012impact} investigate how \acs{WAMS} network failures may bring the system to an unobservable state and cause severe cascading events. In addition, Gu et al.~\cite{gu2015dynamic} propose a dynamic \acs{SE} method that integrates Kalman filtering with Kriging-based forecasting to mitigate communication failures.

\begin{figure*}[ht]
    \centering
    \includegraphics[width=0.9\textwidth]{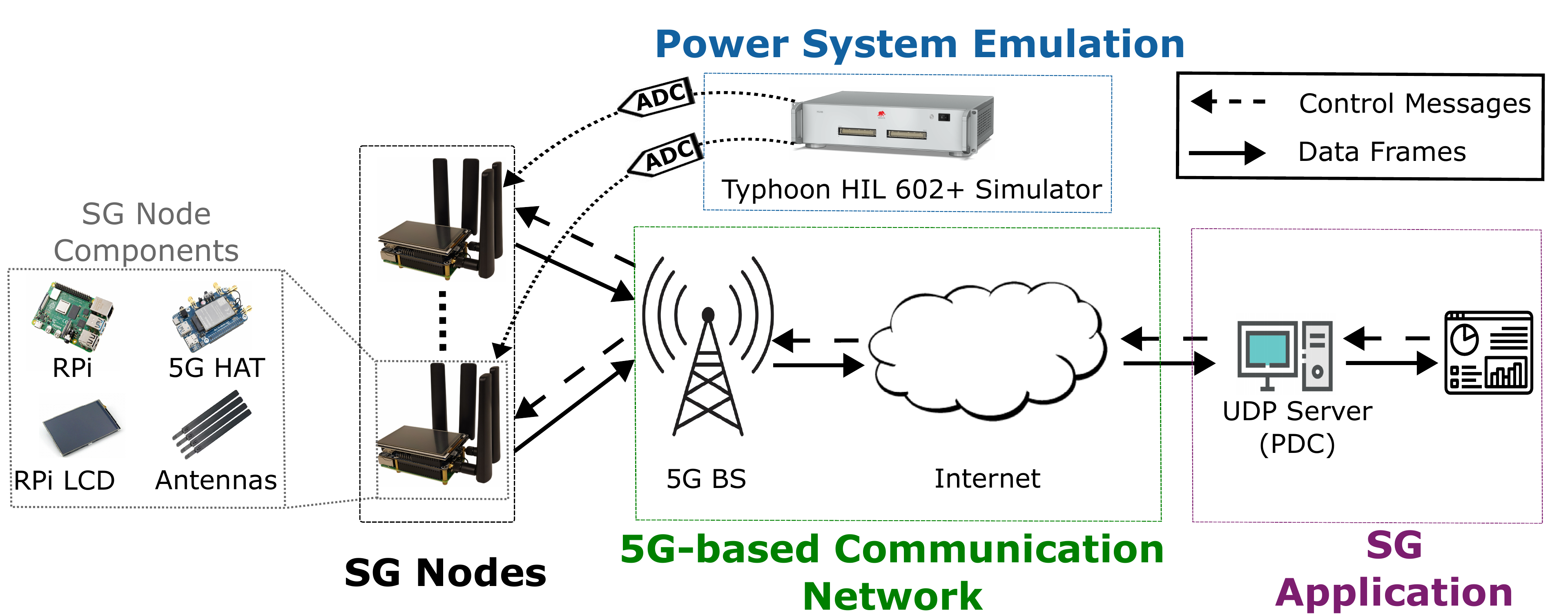}
    \caption{5G-based multi-node Smart Grid (SG) communication testbed using Typhoon HIL simulator.}\label{fig:multiPMU_Typhoon}
\end{figure*}

Despite extensive research on wireless communication technologies for \acp{SG} and increasing interest in 5G-enabled monitoring and estimation, several important gaps remain. First, most existing studies rely on analytical models or offline simulations, thereby lacking experimental validation. While simulations and analytical methods can be cost-effective for predicting communication network behavior during the design phase, experimental studies provide tangible insights into the complexities of \acs{SG} communication systems. Second, existing work rarely provides real-world experimental validation of 5G performance using actual grid measurements. Finally, there is a lack of integrated experimental platforms that combine real hardware-based \acs{SG} nodes, commercial 5G networks, \ac{HIL} power-system simulators, and application-level evaluations, such as \acs{SE}, load tracking, and fault detection. These gaps are addressed in this paper by conducting an experimental campaign on a custom-built testbed over a commercial 5G network. 
\section{System Description} \label{section:system}

Fig.~\ref{fig:multiPMU_Typhoon} illustrates the architecture of our multi-node testbed, which is 
designed as an end-to-end cyber-physical platform consisting of four tightly coupled layers: i) \emph{power system emulation}, 
ii) \emph{SG node}, 
iii) \emph{5G-based communication network}, and 
iv) \emph{SG application}.


The main function of the \emph{power system emulation} layer is to provide realistic grid dynamics. To do so, we used a Typhoon \acs{HIL}~602+ real-time simulator, 
with the Typhoon \acs{HIL} Control Center software running on a host computer. The reference scenario considered in this work is an IEEE 4-node distribution feeder model that generates real-time analog outputs corresponding to bus voltages, currents, and power measurements. These analog signals are routed through a breakout interface to the SG nodes.

Each \emph{SG node} performs computation tasks, using a \acs{RPi}~4 Model~B device, and communication tasks, using an external 5G modem (i.e., Quectel RM520N-GL 5G HAT) and four cellular antennas, as shown in Fig.~\ref{fig:multiPMU_Typhoon}. Each SG node also features an LCD screen for local visualization and operation. The SG nodes interface directly with the Typhoon \acs{HIL} simulator to acquire measurement signals. Since the \acs{RPi} does not include a built-in \ac{ADC}, each node employs an external 16-bit ADS1115 \acs{ADC} to digitize the analog outputs produced by the HIL simulator. To accommodate both positive and negative signal values, measurements are appropriately scaled within Typhoon to match the ADC input range, and inverse scaling is applied at the SG node to reconstruct the original physical quantities. Each SG node locally processes the digitized measurements and packages them into synchrophasor-like data frames compliant with the IEEE C37.118.2 Standard~\cite{IEEE_C37.118.2}. These frames are transmitted at configurable reporting rates over the commercial 5G network.

The \emph{5G-based communication network} layer provides bidirectional connectivity between the \ac{SG} nodes and the \ac{SG} application layer. 
Measurement frames generated by the SG nodes traverse the 5G \ac{BS}, the operator's core network, and the public Internet before reaching the server. Bidirectional connectivity is also used to transmit control messages from the server, enabling dynamic adjustments to parameters such as reporting rate, payload size, and experiment configuration.

The \emph{SG application} layer consists of a UDP server running on a remotely connected virtual machine, which is in charge of exchanging data packets with the SG nodes as well as aggregating the received data and executing the tasks associated with the selected SG application. In the considered synchrophasor-like scenario, the UDP server acts as a \ac{PDC}. In this work, we focus on an \ac{SE} application; however, the proposed testbed can be used to emulate a variety of SG applications, including \ac{PMU}, distributed optimization of the power distribution system, and smart metering~\cite{11204602}: this aspect is presented in 
Section~\ref{sec:DelayAnalysis}. Then, we formulate the real-time \acs{SE} problem and experimentally validate its performance under diverse operating scenarios (Section~\ref{section:SE}).
\section{Experimental Performance Evaluation of 5G for Smart Grid Communications} \label{sec:DelayAnalysis}

To assess the suitability of the commercial 5G network and our custom-built SG nodes for \acs{SG} communications, we conducted a series of experiments designed to emulate realistic \acs{SG} traffic patterns. These experiments evaluate communication performance and are conducted using SG nodes without the Typhoon \acs{HIL} simulator, thereby isolating communication from power system dynamics. Specifically, we study the communication performance of the commercial 5G network over a range of reporting rates (Section~\ref{section:framerates}) and compare its performance under indoor and outdoor deployment scenarios (Section~\ref{section:inVsout}). 

Each SG node was configured to transmit frames \textit{periodically} at a rate of $\lambda$ frames per second (fps), a setup flexible enough to represent various \acs{SG} data types (e.g., synchrophasors, smart metering, and monitoring data). We conducted an end-to-end performance evaluation focusing on three network \acp{KPI}, including delay, jitter, and frame loss. End-to-end delay was computed \textit{at frame level} by subtracting the transmission time from the reception time of each frame and was then used\footnote{To use this method, the clocks of both transmitter and receiver need to be synchronized with high accuracy: this is made possible thanks to the use of the Network Time Protocol (NTP), which ensures a synchronization error in the order of a few ms.}. Here, the delay accounts for the 5G transmission delay and internet-induced latency.

\subsection{5G Communication Performance Under Different Reporting Rates}  \label{section:framerates}

To evaluate the suitability of the commercial 5G network for supporting a broad range of SG applications, we conducted an experimental study over reporting rates representative of those applications. In particular, we focused on the IEEE Standard C37.118.2~\cite{IEEE_C37.118.2}, analyzing power systems operating at two possible frequency, i.e., $50$ Hz (with reporting rate $\lambda = 0.5,1,10, 25, 50$ and $100$ \ac{fps}~\cite{IEEE_C37.118.2}) and $60$ Hz (with $\lambda = 12, 15, 20, 30,60$ and $120$ fps~\cite{IEEE_C37.118.2}). 
For every value of $\lambda$, three experiments were run: each experiment featured the transmission of 30000 frames, except for smaller reporting rates such as 10--15 fps (15000 frames), and 0.5--1 fps (7500 frames). All experiments were conducted using a single \acs{SG} node deployed in an indoor environment (on the University at Buffalo campus). These data were then aggregated for all experiments with the same reporting rate and results were summarized in Table~\ref{table:delay_stats_reportingRate}. For each reporting rate, we considered delay information of all the frames transmitted in the three experiments and computed minimum, maximum, mean, standard deviation (St.d.), first and third quartiles (Q1, Q3), jitter, 95\% Confidence Interval (95\% CI), and frame loss percentage. 
\begin{table}[hbt!] 
\caption{Delay Statistics vs. Reporting Rate.} 
\centering
\setlength{\tabcolsep}{3.5pt} 
\renewcommand{\arraystretch}{1.5} 
\resizebox{\columnwidth}{!}{
\begin{tabular}{c||cccccccc|c}
\multirow{2}{*}{\shortstack[l]{Reporting\\rate $\lambda$ fps}} & \multicolumn{8}{c|}{Delay stats (ms)}  & \multirow{2}{*}{\shortstack[l]{Frame \\ Loss (\%)}}\\
\cline{2-9}
& Min & Max & Mean & St.d. & Q1 & Q3 & Jitter & 95\% CI &\\
\hline\hline
0.5 & 12.59       & 325.73      & 32.66        & 7.77                 & 29.59 & 36.85 & 6.11     & 32.66 $\pm$ 5.37 & 0.00e+00       \\
1 & 16.57       & 170.64      & 32.14        & 3.54                 & 31.00 & 31.48 & 2.00     & 32.14 $\pm$ 3.33 & 0.00e+00       \\

10 & 19.05       & 151.31      & 29.86        & 3.94                 & 27.76 & 31.25 & 1.78     & 29.86 $\pm$ 5.74 & 1.11e-02       \\
12 & 13.59       & 271.97      & 23.58        & 4.72                 & 19.61 & 26.08 & 5.12     & 23.58 $\pm$ 1.47 & 0.00e+00       \\
15 &  13.79       & 192.85      & 22.31        & 5.18                 & 17.81 & 25.02 & 6.83     & 22.31 $\pm$ 2.41 & 4.43e-03       \\
20 & 21.80       & 160.62      & 26.46        & 3.87                 & 24.36 & 27.91 & 2.44     & 26.46 $\pm$ 0.78 & 0.00e+00       \\
25 & 21.00       & 163.46      & 25.87        & 3.95                 & 23.92 & 25.82 & 2.19     & 25.87 $\pm$ 1.08 & 0.00e+00       \\
30 & 16.97       & 144.82      & 23.55        & 3.92                 & 21.12 & 24.76 & 3.46     & 23.55 $\pm$ 1.70 & 0.00e+00       \\
50 & 21.30       & 85.14       & 25.50        & 3.02                 & 24.38 & 24.80 & 2.15     & 25.50 $\pm$ 0.32 & 0.00e+00       \\
60 & 16.50       & 122.15      & 21.99        & 4.33                 & 18.66 & 24.03 & 4.21     & 21.99 $\pm$ 0.89 & 0.00e+00       \\
100 &16.99       & 92.92       & 22.09        & 4.34                 & 19.34 & 24.36 & 4.04     & 22.09 $\pm$ 0.88 & 0.00e+00       \\
120 & 16.59       & 156.50      & 20.44        & 5.11                 & 18.22 & 21.09 & 3.56     & 20.44 $\pm$ 0.44 & 0.00e+00       \\

\end{tabular}}
\label{table:delay_stats_reportingRate}
\end{table}

Across all reporting rates, the 5G-based communication network consistently demonstrates stable and robust performance with a low mean delay (20-33 ms), a low jitter, and frame loss nearing zero across the full range. However, a mild trend is observed at very low reporting rates (0.5--10\,fps), where the delay is slightly higher (29--32\,ms) and more variable. This behavior can be attributed to larger inter-frame intervals, which allow channel conditions to drift and may cause the modem to re-initiate scheduling or random-access procedures. However, these effects are modest, and the performance remains well within acceptable limits for \acs{SG} applications. Once the reporting rate reaches the IEEE-recommended typical operational range (12--60\,fps) and beyond, the system stabilizes. The mean delay converges to 20--26\,ms and 95\% CIs become narrow, indicating that frequent transmissions keep the radio link consistently engaged and support smoother scheduling at the cellular BS. Even at extreme rates of 100--120\,fps---well above conventional \ac{PMU} practice--the 5G link maintains low-latency, low-variability performance with no observable congestion or frame loss.

These results reveal that 5G network performance is largely insensitive to the reporting rates we used, thus making 5G (and our testbed) suitable for a wide variety of SG applications.

\subsection{Indoor vs. Outdoor Communication Performance } \label{section:inVsout}

Depending on the chosen SG application, nodes may be installed indoor or outdoor, we further examined the network communication performance by comparing the differences in these two scenarios: 
we conducted three indoor and three outdoor experiments using a single \acs{SG} node. 
As in Table~\ref{table:delay_stats_reportingRate}, we computed statistics at \textit{frame level} and presented comparative results indoor vs outdoor in Table~\ref{table:delay_stats_multi_pmu}. 

\begin{table}[hbt!]
\caption{{Delay Statistics and Frame Loss in Indoor and Outdoor Experiments.}}
\centering
\setlength{\tabcolsep}{2.5pt} 
\renewcommand{\arraystretch}{1.5} 
\resizebox{\columnwidth}{!}{
\begin{tabular}{c||cccccccc|c}
\multirow{2}{*}{\shortstack[l]{Setup}}  & \multicolumn{8}{c|}{Delay stats (ms)}  & \multirow{2}{*}{\shortstack[l]{Frame \\ Loss (\%)}}\\
\cline{2-9}
& Min & Max & Mean & St.d. & Q1 & Q3 & Jitter & 95\% CI &\\
\hline\hline
Indoor & 21.30       & 85.14       & 25.50        & 3.02                 & 24.38 & 24.80 & 2.15     & 25.50 $\pm$ 0.32 & 2.22e-03 \\

\hline 
Outdoor &  15.41       & 124.00      & 22.35        & 5.05                 & 18.09 & 24.50 & 4.05     & 22.35 $\pm$ 1.82 & 1.11e-03       \\

\end{tabular}}
\label{table:delay_stats_multi_pmu}
\end{table}

Both indoor and outdoor experiments exhibit very similar, highly reliable, and low-latency 5G performance. The mean delay ranges from 22.35\,ms (outdoor) to 25.50\,ms (indoor), a substantial improvement over our previous LTE cat-M system, which reported delays of approximately 176\,ms outdoors and 200\,ms indoors~\cite{10333918}. Jitter is also low in both settings, ranging from 2.15\,ms (indoor) to 4.05\,ms (outdoor), indicating stable transmission. The St.d. and 95\% CI remain close, and the frame loss rate is in the order of $10^{-3}$ (which is considered very low for most of the envisioned SG applications).

\begin{figure}
    \centering
    \includegraphics[scale=0.14]{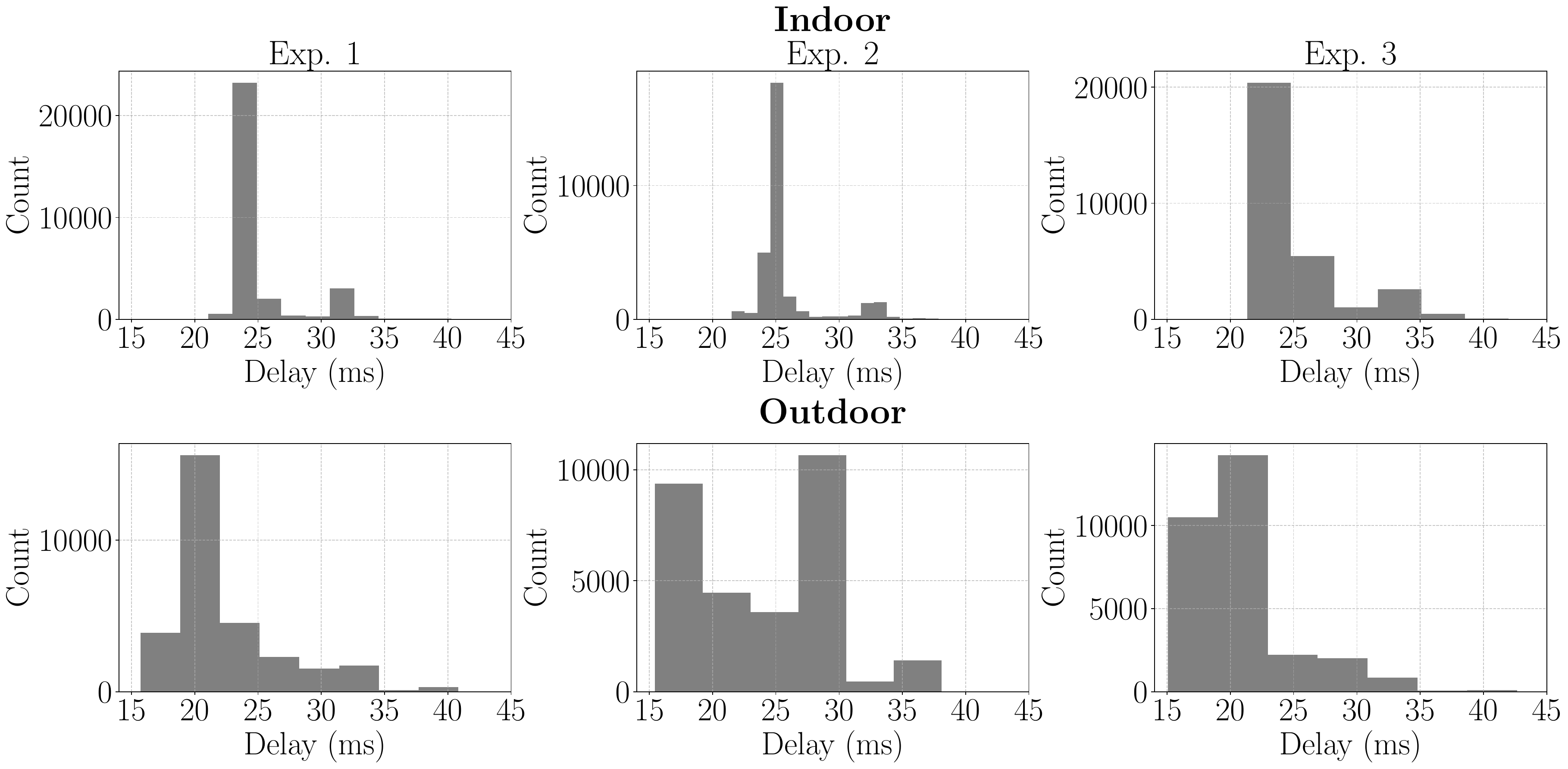}
    \caption{Comparison of delay distributions in indoor and outdoor environments.}    
    \label{fig:Frameloss_hist}
\end{figure}

Although the performance is similar, we observe two subtle differences: (1) indoor experiments show a slightly higher mean delay than the outdoor experiments, and (2) outdoor experiments exhibit a broader delay spread (higher St.d. and jitter) than the indoor experiments. These differences become more apparent in the delay histograms in Fig.~\ref{fig:Frameloss_hist} across all three experiments. For visualization clarity, the delay distributions are truncated at the $99.9^\text{th}$ percentile to remove rare extreme outliers (<0.1\% of samples). All reported metrics in Table~\ref{table:delay_stats_multi_pmu} are computed using the full dataset.

In indoor experiments, the majority of delay samples cluster very tightly around 24--28\,ms, with only small tails extending beyond 30\,ms. In contrast, outdoor delays span a wider range: a large number of samples fall between 16--20\,ms, and smaller clusters appear around 24--30\,ms. This pattern explains why outdoor experiments have a slightly lower mean delay (because of the strong concentration of delays below 20\,ms) but a larger overall variation. This pattern can be explained by the scenarios under which the experiments were conducted. The outdoor experiments took place during winter in Buffalo, NY, at approximately 0$\degree$C, near a UB parking lot during late afternoon, coinciding with typical end-of-office hours, when there was continuous movement of cars and pedestrians. Such mobility may introduce time-varying multipath reflections and intermittent micro-blockages, while low ambient temperatures may affect hardware performance. Collectively, these factors result in links that are generally faster (benefiting from line-of-sight propagation) but more dynamic, leading to increased variability in delay. By contrast, the indoor environment provided controlled temperature and minimal movement, resulting in a more static propagation setting. Although indoor propagation experiences additional penetration losses, which slightly elevate the average delay, the channel conditions are more stable over time, leading to lower jitter, smaller St.d., and a tighter Q1-Q3 range. Thus, indoor performance is slightly slower but more stable, whereas outdoor performance is slightly faster but more variable.

Across all experimental results presented in Section~\ref{sec:DelayAnalysis}, the highest mean delay observed for the 5G was 32.66\,ms, occurring at a reporting rate of 0.5\,fps (Section~\ref{section:framerates}). At the same reporting rate, our previous LTE cat-M study reported a mean delay of 212.52\,ms~\cite{10333918}. Thus, even under the worst-case delay observed for the 5G system, the 5G achieves approximately 6.5$\times$ lower mean delay than LTE cat-M.

\section{Real-time State Estimation Framework and Experimental Validation over 5G Networks}  \label{section:SE}

\acs{SE} is a crucial analytical process in modern power systems that reconstructs the grid's operational state by estimating voltage magnitudes and angles from available measurements. Accurate \acs{SE} enhances system observability, supports load forecasting, improves stability assessment, and is frequently used as a foundation for event detection and control actions.

In Section~\ref{sec:DelayAnalysis}, we evaluated the communication performance of the proposed 5G-enabled \acs{SG} nodes using simulated data streams that emulate synchrophasor traffic. While those experiments demonstrated the suitability of 5G for low-latency and lightweight \acs{SG} communications, they did not incorporate realistic power system dynamics. In this section, we extend the analysis to a closed-loop cyber-physical setting by experimentally evaluating real-time \acs{SE} using synchrophasor measurements obtained from an IEEE 4-node distribution feeder model implemented on a Typhoon \ac{HIL} real-time simulator.

This section first describes the real-time \acs{SE} framework implemented using the proposed testbed, including the power system model and the mathematical formulation of the \acs{SE} problem (Section~\ref{subsection:Realtime_SE}). It then presents an experimental validation of the \acs{SE} framework under diverse operating conditions, including steady-state operation, dynamic load variations, and fault events (Section~\ref{subsection:SEResults}).

\subsection{Real-Time SE Framework} \label{subsection:Realtime_SE}

\subsubsection{Power System Model} \label{subsection:PowerSystem}

\begin{figure}
    \centering
    \includegraphics[scale=0.6]{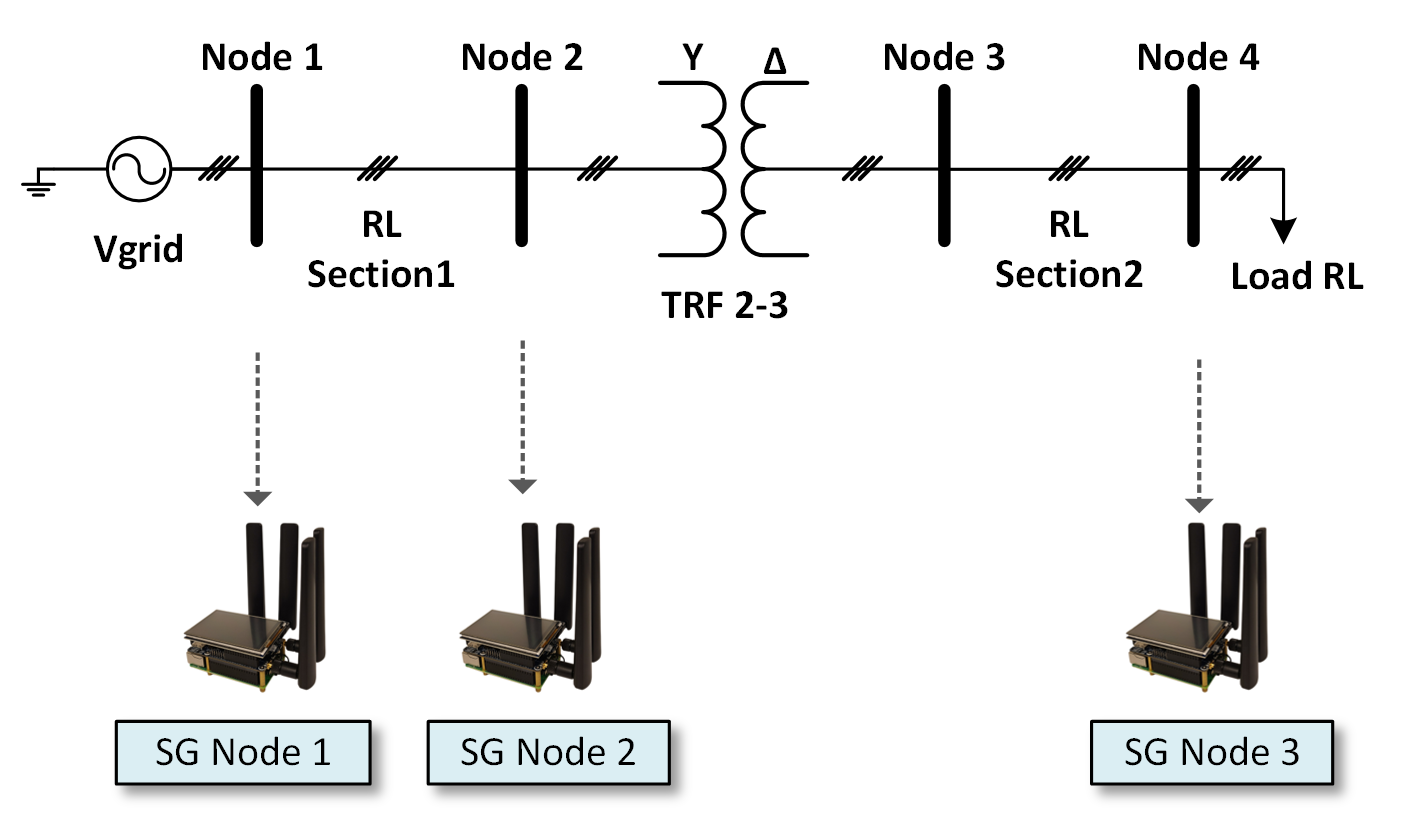}
    \caption{IEEE 4-node power system model.}
    \label{fig:power_system}
\end{figure}

The experiments were conducted using the testbed shown in Fig.~\ref{fig:multiPMU_Typhoon}, which is based on an IEEE 4-node distribution feeder model implemented on a Typhoon HIL 602+ real-time simulator. The feeder, shown in Fig.~\ref{fig:power_system}, consists of four buses interconnected through resistive-inductive (RL) line segments, capturing realistic voltage drops and phase-angle variations along the feeder. Power grid Node 1 interfaces with an external grid through a reference voltage source, setting the system voltage and frequency. Power flows from Node~1 to Node~2 via an RL line, from Node~2 to Node~3 through a distribution transformer, and from Node~3 to Node~4 through another RL segment.

For this study, measurements were collected at Nodes~1, 2, and~4 using three \acs{RPi}-based \acs{SG} nodes as shown in Fig.~\ref{fig:power_system}. At Node 1, the measured quantities included voltage magnitude, current magnitude, and active power. At Node 2, voltage magnitude, voltage angle, and current magnitude were measured. At Node 4, voltage magnitude, voltage angle, and active power were acquired. All remaining electrical states in the system were treated as unknown and were estimated by the \acs{SE} algorithm.

\subsubsection{Mathematical SE Framework} \label{subsection:SE_algorithm}


We adopted the standard \ac{WLS} formulation for the \acs{SE}~\cite{abur2004power}. Let
\begin{equation}
x = [V_1,\, V_2,\, V_3,\, V_4,\, \theta_2,\, \theta_3,\, \theta_4]^\top,
\label{eq:state_vector}
\end{equation}
denotes the system state vector, where \(V_i\) and \(\theta_i\) represent the voltage magnitude and phase angle, respectively, at Node \(i\). Because Node~1 is the reference, its angle is fixed at \(\theta_1=0\) and is omitted from the optimization variables.

At each reporting instant $k$, the measurement vector $z$ consists of all synchrophasor quantities provided by the SG nodes:
\begin{equation}
z = [\hat{V}_1,\, \hat{I}_1,\, \hat{S}_1,\, \hat{V}_2,\, \hat{\theta}_2,\, \hat{I}_2,\, \hat{V}_4,\, \hat{\theta}_4,\, \hat{S}_4]^{\top},
\label{eq:measurement_vector}
\end{equation}
where $\hat{(\cdot)}$ denotes a real-time measurement. All quantities were converted to per-unit before processing.

Each measurement $z_i$ is related to the system states through a nonlinear function $h_i(x)$~\cite{abur2004power}:
\begin{equation}
z_i = h_i(x) + \varepsilon_i,
\end{equation}
where $\varepsilon_i$ represents measurement noise. The complete measurement model is written compactly as
\begin{equation}
z = h(x) + \varepsilon.
\end{equation}

\textit{WLS SE Problem:} The estimator computes the state $x$ by minimizing the weighted squared residual~\cite{abur2004power}:
\begin{align}
\begin{array}{l}
     \min_{x} \; F(x)
        = \sum_{i=1}^{m}
        w_i \left[z_i - h_i(x)\right]^2
        = \| r(x) \|_W^2, \\
        \text{s.t.} \\
        c_i(x) = 0,\;\;\;\text{for } i\in\mathcal{N}_{int}        \\
        x\in\mathcal{X}
\end{array}
\end{align}
where $r(x) = z - h(x)$ is the residual vector, $c_i(x)$ are the power flow functions at internal nodes $\mathcal{N}_{int}$ (e.g. Nodes 2 and 3 in Fig. \ref{fig:power_system}), $\mathcal{X}$ is a convex set used to bound the states (typically upper and lower bounds), and $W = \mathrm{diag}(w_1,\dots,w_m)$ is the diagonal weighting matrix.

The optimization enforces the AC power-balance equations at the interior nodes (Node~2 and Node~3), together with reasonable operating bounds on voltage magnitudes (0.5--1.0~p.u.) and phase angles (\(-70^\circ\) to \(0^\circ\)). The problem is solved using MATLAB's \texttt{fmincon} optimization solver with the Sequential Quadratic Programming (SQP) algorithm, which iteratively updates $x$ using the nonlinear measurement model and its Jacobian.

\textit{Residual Norm and Fault Indicator:} After each SE update at reporting instant $k$, the measurement residual is computed as
\begin{equation}
\label{eq:residual}
r(k) = z_{\text{meas}}(k) - z_{\text{est}}(k),
\end{equation}
where $z_{\text{est}}(k) = h(x^\ast(k))$ is the predicted measurement at the converged estimate $x^\ast(k)$ and $z_{\text{meas}}(k)$ is the measured value. The corresponding \acs{WLS} cost function value is
\begin{equation}
\label{eq:costfn}
F(k) = \| r(k) \|_W^2.
\end{equation}

The scalar value \(F(k)\) provides a global measure of mismatch between the system model and the measured data and is used as a fault-detection metric in this study. Under normal operating conditions, \(F(k)\) remains below a threshold value, while abrupt disturbances cause sharp increases. A fault is declared whenever

\begin{equation}
\label{eq:costfn_thrshold}
F(k) > T,
\end{equation}
where $T$ is a detection threshold selected based on the statistical behavior of $F(k)$ under normal operating conditions. 

\subsection{Experimental Validation of SE Under Diverse Operating Conditions} \label{subsection:SEResults}

To evaluate the \acs{SE} performance, the measured synchrophasor signals are compared with their corresponding model-based estimates at each reporting instant $k$. The frame ID serves as the reporting instant \(k\). For each frame, we compute the absolute residual from~(\ref{eq:residual}), which captures the instantaneous mismatch between measured and estimated quantities. To quantify overall estimation accuracy across all frames, we use the \ac{MAE} and \ac{RMSE}, defined as

\begin{equation}
\label{RMSD}
\begin{aligned}
\mathrm{MAE} \;&=\; \frac{1}{n} \sum_{k=1}^{n} |z_{\text{meas}}(k) - z_{\text{est}}(k)|, \\
\mathrm{RMSE} \;&=\; \sqrt{\frac{1}{n} \sum_{k=1}^{n} \left( z_{\text{meas}}(k) - z_{\text{est}}(k) \right)^2},
\end{aligned}
\end{equation}
where $z_{\text{meas}}(k)$ and $z_{\text{est}}(k)$ denote the measured and estimated values of a given signal at frame ID \(k\), and \(n\) is the total number of frames.

To evaluate the performance of the \acs{SE} framework, three experimental scenarios are considered:
\begin{itemize}
    \item \textbf{SE under steady-state conditions}, which assesses estimation accuracy during stable operation;
    \item \textbf{SE under dynamic load conditions}, which evaluates \acs{SE} performance under controlled load variations;
    \item \textbf{Fault detection under stable load conditions}, which examines the ability of the \acs{SE} framework to detect abrupt disturbances.
\end{itemize}

All experiments were conducted indoors using the proposed testbed. Each \acs{SG} node transmits 30000 synchrophasor frames over the commercial 5G network. A reporting rate of 50\,fps was used for the steady-state condition and fault-detection experiments, while 10\,fps was used for dynamic load scenarios.

\subsubsection{SE Under Steady-State Conditions} \label{subSec:baselineSE}

\begin{table}[t]
    \centering
    \caption{Communication Performance of \acs{SG} Nodes Over the Baseline SE Experiment.}
    \setlength{\tabcolsep}{3.5 pt}
    \begin{tabular}{|c|c|c|c|c|c|c|}
    \hline
    SG node  & Sent & Received & Lost & Loss & Mean & Jitter \\
    location per bus & frames & frames & frames & rate (\%) & delay (ms) & (ms) \\
    \hline \hline
    Node 1 & 30000 & 30000 & 0 & 0.00 & 25.38 & 3.17 \\ 
    Node 2 & 30000 & 30000 & 0 & 0.00 & 31.93 & 3.02 \\ 
    Node 4 & 30000 & 30000 & 0 & 0.00 & 27.54 & 3.36 \\ 
    \hline
    \end{tabular}
    \label{tab:SE_Com_RPI}
\end{table}

This experiment serves as the baseline case for evaluating the intrinsic accuracy of the \acs{SE} framework under stable operating conditions. No load variations or faults were introduced in this experiment.

Table~\ref{tab:SE_Com_RPI} summarizes the end-to-end communication performance during data collection. All three nodes experienced zero frame loss, indicating a reliable 5G link during the experiment. The average delay ranged from 25 to 32\,ms, and jitter remained low (approximately 3\,ms across all nodes), consistent with the results reported in Section~\ref{sec:DelayAnalysis}. These stable network conditions ensure that the observed \acs{SE} performance reflects the electrical behavior of the power system model rather than communication-induced effects.

\begin{table}[t]
	\centering
    \caption{Baseline SE Accuracy Across the Nodes.}
        \setlength{\tabcolsep}{5pt}
        \begin{tabular}{|c|c|c|}
	\hline
	Signal type & {\acs{MAE}} & {\acs{RMSE}}\\ 
        \hline \hline        
        Node 1 $Vmag$ (V) & 0.68 & 0.78 \\ 
        Node 2 $Vmag$ (V) & 0.65 & 0.69\\ 
        Node 2 $Vang$ (deg) & 0.07 & 0.07 \\ 
        Node 4 $Vmag$ (V) & 0.58 & 0.74 \\ 
        Node 4 $Vang$ (deg) & 0.56 & 0.69 \\ 
        \hline  
	\end{tabular}
	\label{tab:SE_Accuracy_RPI}
\end{table}

For each subfigure, the measured and estimated results are shown at the top, and the corresponding absolute residual errors are shown at the bottom. The associated \acs{MAE} and \acs{RMSE} values are reported in Table~\ref{tab:SE_Accuracy_RPI}.

\begin{figure}
    \centering
    \includegraphics[width=1\linewidth]{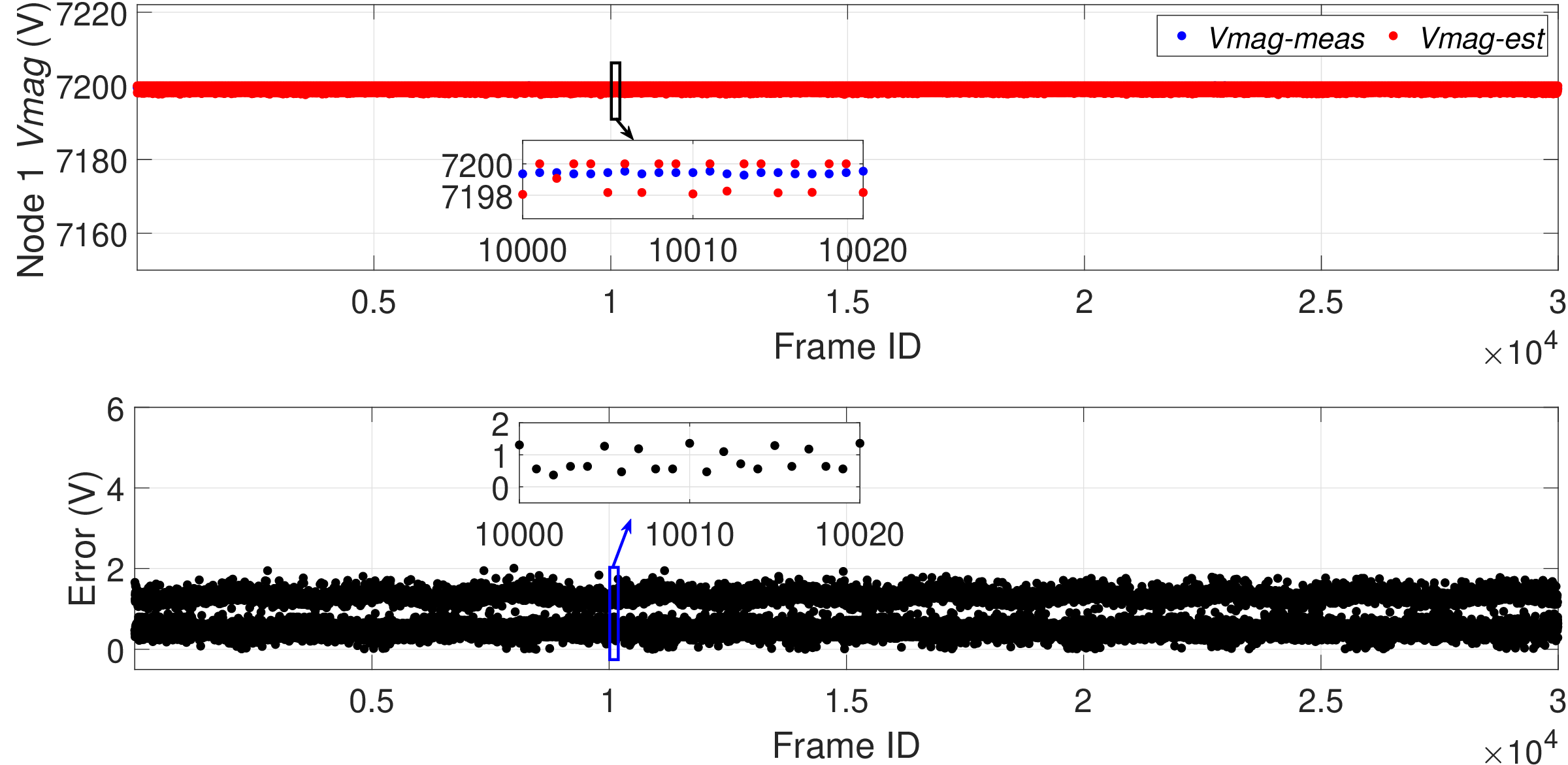}
    \caption{Baseline SE at Node~1: measured vs.\ estimated $V_{mag}$ with residual errors.}
    \label{fig:Node1SE_P}
\end{figure}

Fig.~\ref{fig:Node1SE_P} shows the measured voltage magnitude ($Vmag\text{-}meas$) and estimated voltage magnitude ($Vmag\text{-}est$) over the total number of frames for Node~1. The results indicate that $Vmag\text{-}est$ closely aligns with $Vmag\text{-}meas$, as observed in both the estimation plot and the error plot. The residual error remains consistently small across all frames, with no noticeable deviations. The \acs{MAE} and \acs{RMSE} are approximately 0.68\,V and 0.78\,V, respectively, confirming strong estimation accuracy. This shows that the \acs{SE} algorithm effectively estimated the measured voltage magnitude for Node 1.


\begin{figure}[!tbp]
  \centering
  \begin{minipage}{.475\textwidth}
      \centering
      \includegraphics[width=\linewidth]{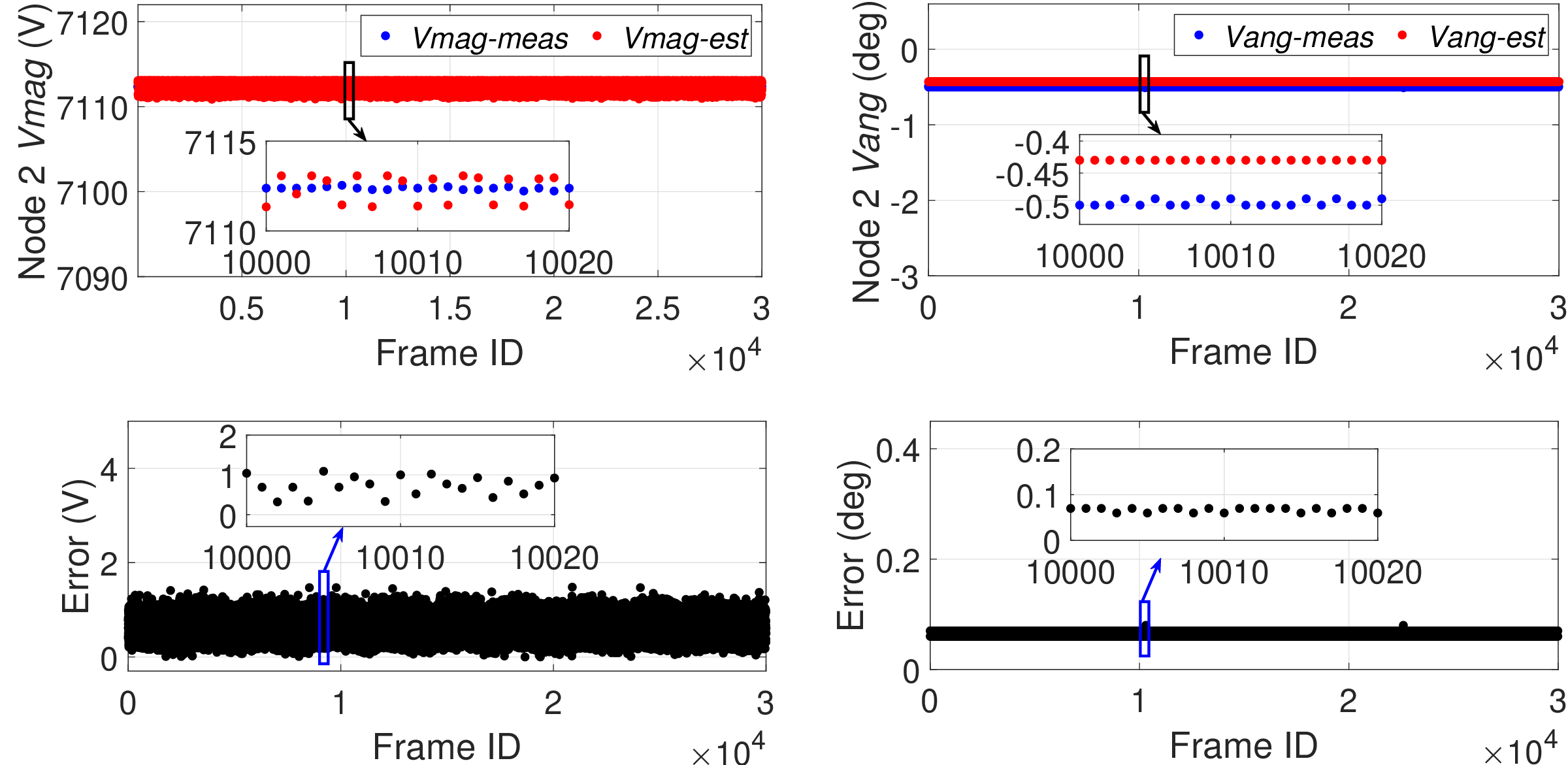}
      \begin{minipage}{.5\linewidth}
        \caption*{\small (a) $Vmag$: meas vs. est}
      \end{minipage}%
      \begin{minipage}{.5\linewidth}
        \caption*{\small (b) $Vang$: meas vs. est}
      \end{minipage}
    \end{minipage}
    \caption{Baseline SE at Node~2: measured vs.\ estimated (a) $V_{mag}$ and (b) $V_{ang}$ with residual errors.}
  \label{fig:Node2SE_P}
\end{figure}
 
Fig.~\ref{fig:Node2SE_P} shows the voltage magnitude ($Vmag$) and angle ($Vang$) estimation results for Node 2. As shown in Fig.~\ref{fig:Node2SE_P}(a), $Vmag\text{-}est$ closely follows $Vmag\text{-}meas$. The error subplot shows that the residuals remain low and stable across the frames. The resulting \acs{MAE} and \acs{RMSE} for $Vmag$ are 0.65\,V and 0.69\,V, respectively. Similarly, the voltage angle results in Fig.~\ref{fig:Node2SE_P}(b) shows strong agreement between estimated ($Vang\text{-}est$) and the measured values ($Vang\text{-}meas$). The residuals remain confined within a narrow band, and both the \acs{MAE} and \acs{RMSE} values are 0.07 deg.

\begin{figure}[!tbp]
  \centering
  \begin{minipage}{.475\textwidth}
      \centering
      \includegraphics[width=\linewidth]{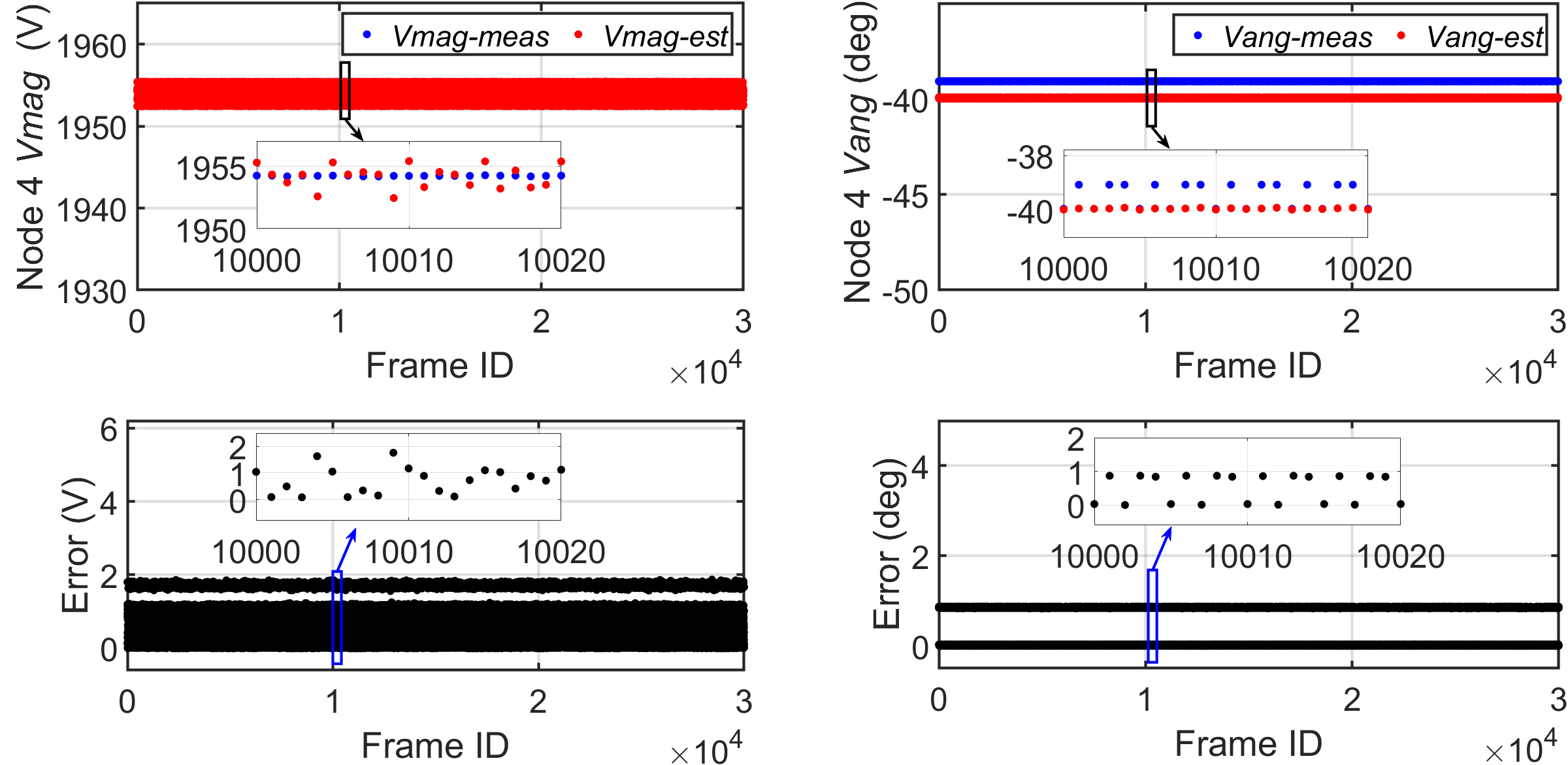}
      \begin{minipage}{.5\linewidth}
        \caption*{\small (a) $Vmag$: meas vs. est}
      \end{minipage}%
      \begin{minipage}{.5\linewidth}
        \caption*{\small (b) $Vang$: meas vs. est}
      \end{minipage}
    \end{minipage}
    \caption{Baseline SE at Node~4: measured vs.\ estimated (a) $V_{mag}$ and (b) $V_{ang}$ with residual errors.}
  \label{fig:Node4SE_P}
\end{figure}

Similar trends are observed for Node~4 in Fig.~\ref{fig:Node4SE_P}, where both $Vmag\text{-}est$ and $Vang\text{-}est$ closely track their measured counterparts. The corresponding \acs{MAE} and \acs{RMSE} values are 0.58\,V and 0.74\,V for the $Vmag$, and 0.56\,deg and 0.69\,deg for $Vang$, confirming consistently accurate estimation performance.

\subsubsection{SE Under Dynamic Load Conditions} \label{subsec:loadtrack}

\begin{table}[t]
    \centering
    \caption{Communication Performance of \acs{SG} Nodes During SE Experiment Under Dynamic Load Conditions.}
    \setlength{\tabcolsep}{3.5 pt}
    \begin{tabular}{|c|c|c|c|c|c|c|}
    \hline
    SG node  & Sent & Received & Lost & Loss & Mean & Jitter \\
    location per bus     & frames & frames & frames & rate (\%) & delay (ms) & (ms) \\
    \hline \hline
	Node 1 & 30000 & 29999 & 1 & 0.01 & 37.13 & 2.17 \\
	Node 2 & 30000 & 30000 & 0 & 0.00 & 33.57  & 1.60 \\
	Node 4 & 30000 & 30000 & 0 & 0.00 & 30.69 & 0.69 \\ 
    \hline
    \end{tabular}
	\label{tab:LC_Com}
\end{table}

This experiment evaluates the ability of the \acs{SE} framework to track time-varying system states under controlled load variations. A periodic square-wave load disturbance was applied at Node~4, switching every 100\,s with a 50\% duty cycle. 

Table~\ref{tab:LC_Com} reports the communication performance during this experiment. Only a single lost frame was observed at Node~1 (0.01\%), while the mean delay ranged from 30.69\,ms to 37.13\,ms. Jitter remained below 2.2\,ms across all nodes. These results indicate stable 5G communication, ensuring that the observed variations in the estimated states are driven by applied load changes rather than by communication effects.

\begin{figure}
    \centering
    \includegraphics[width=1\linewidth]{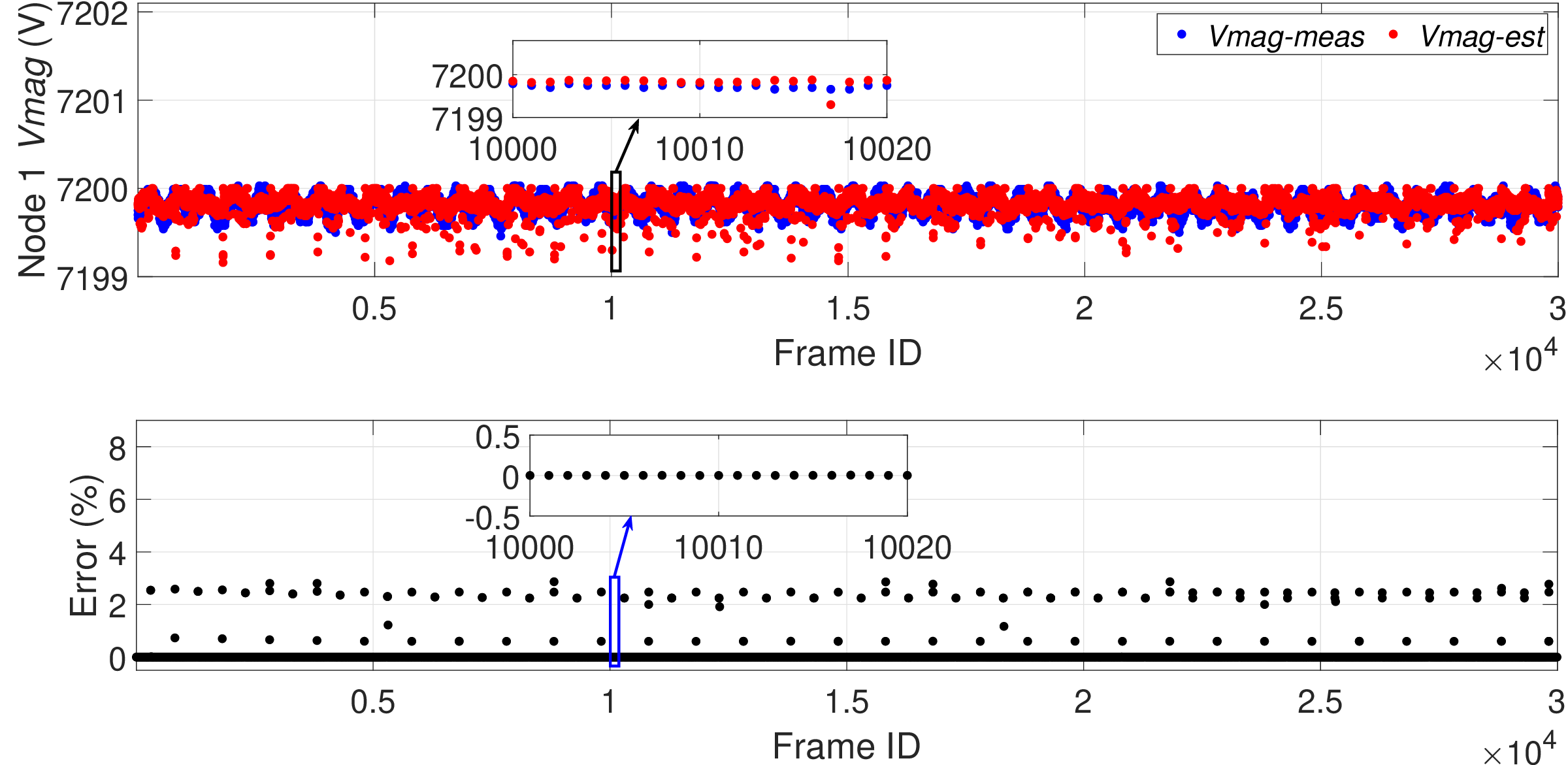}
    \caption{SE performance at Node 1 during dynamic load conditions: measured vs.\ estimated $V_{mag}$ with residual errors.}
    \label{fig:Node1LC_P}
\end{figure}

The Node~1 results in Fig.~\ref{fig:Node1LC_P} show that $Vmag\text{-}est$ generally follows $Vmag\text{-}meas$. The periodic load change every 100\,s is not clearly visible at this node, likely because Node~1 is electrically close to the source in the test feeder and therefore less sensitive to downstream load variations. Although the load change is not directly observed in the measured voltage trajectory, slightly elevated estimation errors appear around the transition instants, as shown by the absolute percentage residuals in Fig.~\ref{fig:Node1LC_P}. This behavior is expected, as the \acs{SE} algorithm jointly computes the system state using all network measurements, causing transient mismatches at nodes unaffected by the load change itself. This effect is reflected in the \acs{MAE} (2.89\,V) and \acs{RMSE} (21.42\,V) in Table~\ref{tab:SE_LoadChange}, where a few large deviations near the switching instants disproportionately raise the \acs{RMSE}. The influence of load variations is more pronounced at Nodes~2 and~4.

\begin{table}[t]
	\centering
 	\caption{SE Accuracy During Load Variations.} 
        \setlength{\tabcolsep}{5pt}
        \begin{tabular}{|c|c|c|}
	\hline
	Signal type & \acs{MAE} & \acs{RMSE}\\ 
        \hline \hline     
        Node 1 $Vmag$ (V) & 2.89 & 21.42 \\ 
        Node 2 $Vmag$ (V) & 2.99 & 20.75 \\
        Node 2 $Vang$ (deg) & 0.06 & 0.06 \\ 
        Node 4 $Vmag$ (V) & 2.81 & 17.40 \\
        Node 4 $Vang$ (deg) & 0.13 & 0.72 \\ 
        \hline  
	\end{tabular}
	\label{tab:SE_LoadChange}
\end{table}

\begin{figure}[!tbp]
  \centering
  \begin{minipage}{.475\textwidth}
      \centering
      \includegraphics[width=\linewidth]{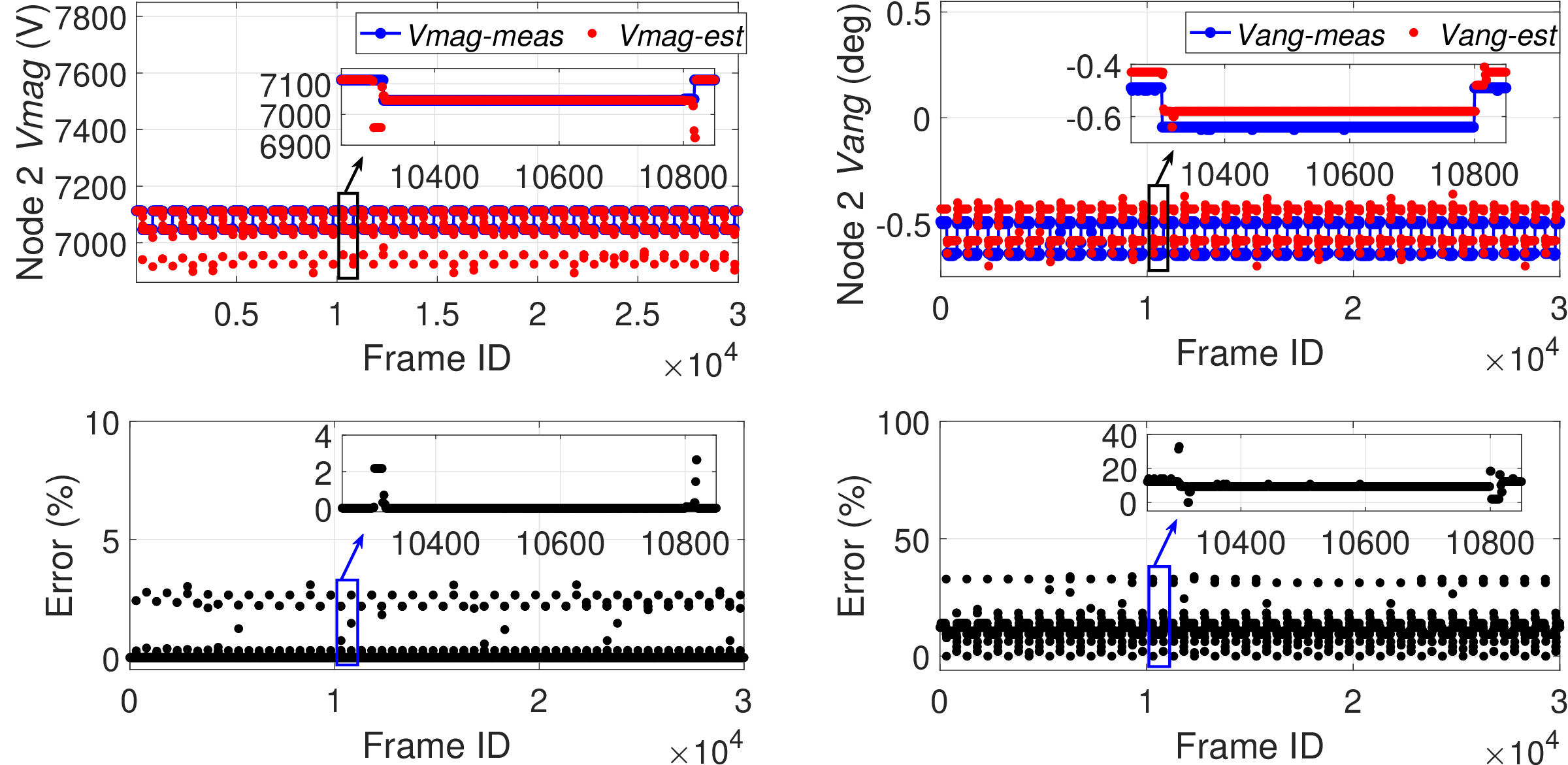}
      \begin{minipage}{.5\linewidth}
        \caption*{\small (a) $Vmag$: meas vs. est}
      \end{minipage}%
      \begin{minipage}{.5\linewidth}
        \caption*{\small (b) $Vang$: meas vs. est}
      \end{minipage}
    \end{minipage}
    \caption{SE performance at Node 2 during dynamic load conditions: measured vs.\ estimated (a) $V_{mag}$ and (b) $V_{ang}$ with residual errors.}
  \label{fig:Node2LC_P}
\end{figure}

The Node~2 results in Fig.~\ref{fig:Node2LC_P}(a) show that $Vmag\text{-}est$ follows the general trajectory of $Vmag\text{-}meas$ and successfully captures the periodic load variations occurring every 100\,s. However, several notable deviations occur around the load-switching intervals, as highlighted in the residual error subplot. These points exhibit error levels of approximately 2--4\%, corresponding to brief misalignment between the measured and estimated values during the state changes. The zoomed-in region of Fig.~\ref{fig:Node2LC_P}(a) confirms that the algorithm momentarily misestimates a small number of samples immediately following each load transition but quickly converges back to the correct trend. The resulting \acs{MAE} and \acs{RMSE} for $Vmag$ are 2.92\,V and 20.41\,V, respectively; the elevated \acs{RMSE} again reflects the influence of a small number of high-error samples during switching events, consistent with the behavior observed at Node~1.

The $Vang$ results in Fig.~\ref{fig:Node2LC_P}(b) show a similar trend. The $Vang\text{-}est$ closely follows the $Vang\text{-}meas$ across the experiment. The residual error plot (bottom) of Fig.\ref{fig:Node2LC_P}(b) shows visually noticeable deviations, but this is primarily due to the small range of measured values (approximately -0.65 deg to -0.45 deg). As a result, even a minor estimation offset---for example, estimating -0.64 deg instead of -0.65 deg--- appears visually amplified when plotted over such a narrow range. The quantitative results confirm high accuracy, with an \acs{MAE} of 0.06\,deg and an \acs{RMSE} of 0.06\,deg, demonstrating the algorithm's ability to reliably estimate $Vang$ at Node~2 even under dynamic load variations.

\begin{figure}[!tbp]
  \centering
  \begin{minipage}{.475\textwidth}
      \centering
      \includegraphics[width=\linewidth]{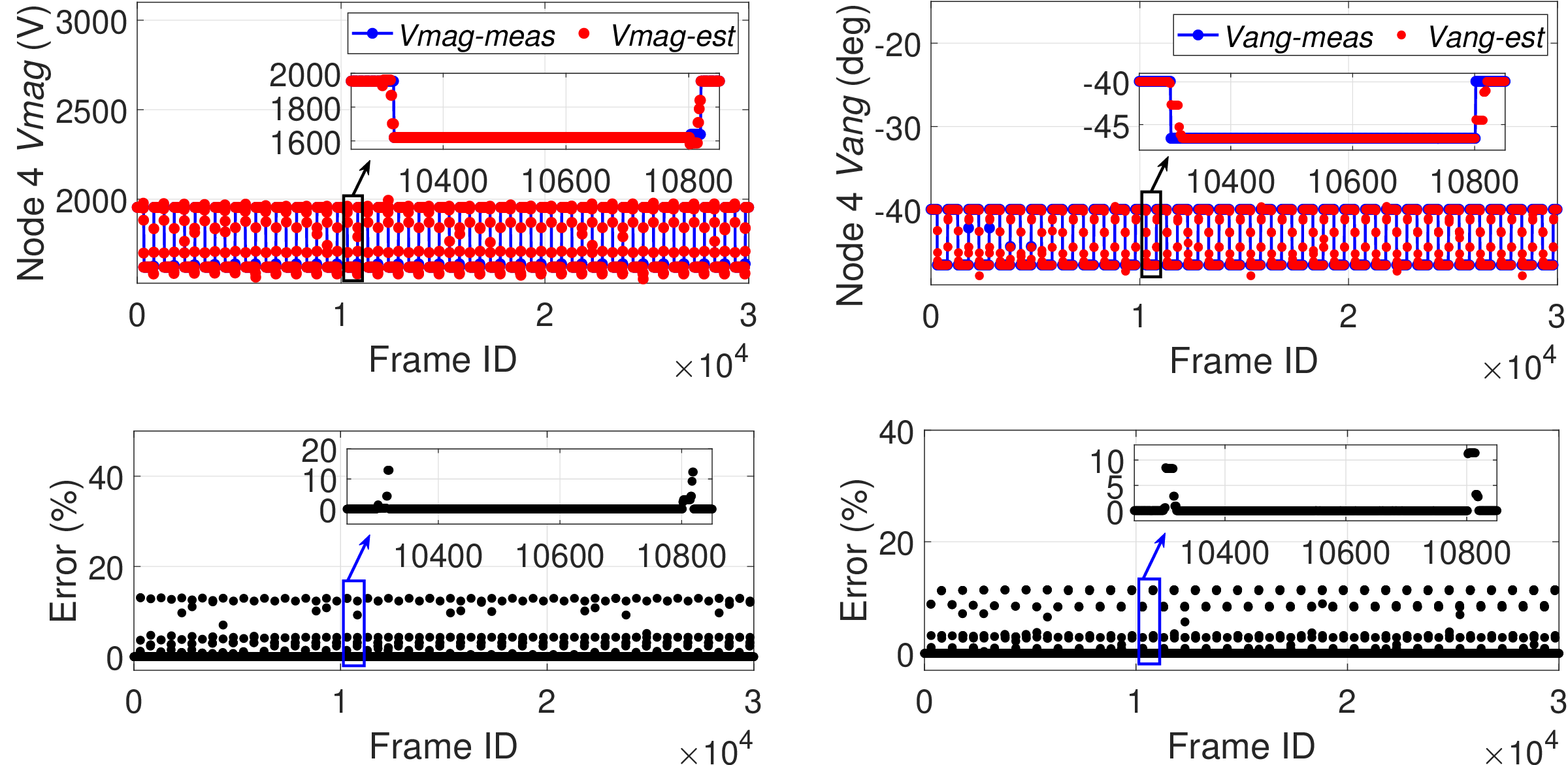}
      \begin{minipage}{.5\linewidth}
        \caption*{\small (a) $Vmag$: meas vs. est}
      \end{minipage}%
      \begin{minipage}{.5\linewidth}
        \caption*{\small (b) $Vang$: meas vs. est}
      \end{minipage}
    \end{minipage}
    \caption{SE performance at Node 4 during dynamic load conditions: measured vs.\ estimated (a) $V_{mag}$ and (b) $V_{ang}$ with residual errors.}
  \label{fig:Node4LC_P}
\end{figure}

Similar behavior is observed for Node~4, as shown in Fig.~\ref{fig:Node4LC_P}. The $Vmag\text{-}est$ generally follows $Vmag\text{-}meas$, with deviations occurring primarily at the load-switching points. The zoomed-in view shows a few estimated samples that briefly overshoot or anticipate the transition before the estimator recovers. The corresponding \acs{MAE} and \acs{RMSE} values for $Vmag$ are 2.34\,V and 17.57\,V, respectively, again reflecting the influence of a small number of high-error samples during abrupt state changes. The $Vang$ results exhibit a similar pattern, with small deviations during transitions and overall low error (\acs{MAE} = 0.13\,deg, \acs{RMSE} = 0.70\,deg). 

These results confirm that the estimator is generally accurate but exhibits brief misalignment during abrupt load transitions. As the downstream-most bus, Node~4 experiences the most pronounced effects of the switching events; nevertheless, the algorithm successfully tracks the system state, with only minor misestimations at the transition points.

\subsubsection{Fault Detection Under Stable Load Conditions} \label{subsec:faultdetection}

This experiment evaluates the robustness of the \acs{SE} framework in detecting abrupt electrical disturbances. Fault events were introduced into the IEEE-4 node feeder model by inserting a fault path between Nodes 3 and 4 via a controllable three-phase switch. Activating the switch connects an additional impedance load to the feeder, producing fault-like behavior characterized by sharp current increases, voltage magnitude drops, and abrupt phase-angle deviations. 

Fault detection is performed using the \acs{WLS} cost function value $F(k)$, which provides a global measure of mismatch between the system model and the measured data. A fault is declared whenever $F(k)$ exceeds a predefined threshold $T$, as defined in~(\ref{eq:costfn}) and~(\ref{eq:costfn_thrshold}). The threshold is selected empirically based on the distribution of $F(k)$ under normal operating conditions and is approximately $6 \times 10^{-3}$.

Table~\ref{tab:FE_Com} summarizes the communication performance during the fault-detection experiment. All nodes maintained reliable connectivity with zero frame loss. The average delay remained close to 30\,ms, and jitter was approximately 3\,ms, indicating a stable communication throughout the experiment.

\begin{table}[t]
    \centering
    \caption{Communication Performance of \acs{SG} Nodes During Fault Detection Experiment.}
    \setlength{\tabcolsep}{3 pt}
    \begin{tabular}{|c|c|c|c|c|c|c|}
    \hline
    SG node  & Sent & Received & Lost & Loss & Mean & Jitter \\
    location per bus & frames & frames & frames & rate (\%) & delay (ms) & (ms) \\
    \hline \hline
	Node 1  & 30000 & 30000 & 0 & 0.00 & 31.28 & 3.39 \\ 
	Node 2 & 30000 & 30000 & 0 & 0.00 & 30.70  & 2.82 \\
	Node 4 & 30000 & 30000 & 0 & 0.00 & 28.86 & 2.94 \\
    \hline
    \end{tabular}
	\label{tab:FE_Com}
\end{table}

\begin{figure}
    \centering
    \includegraphics[width=1\linewidth]{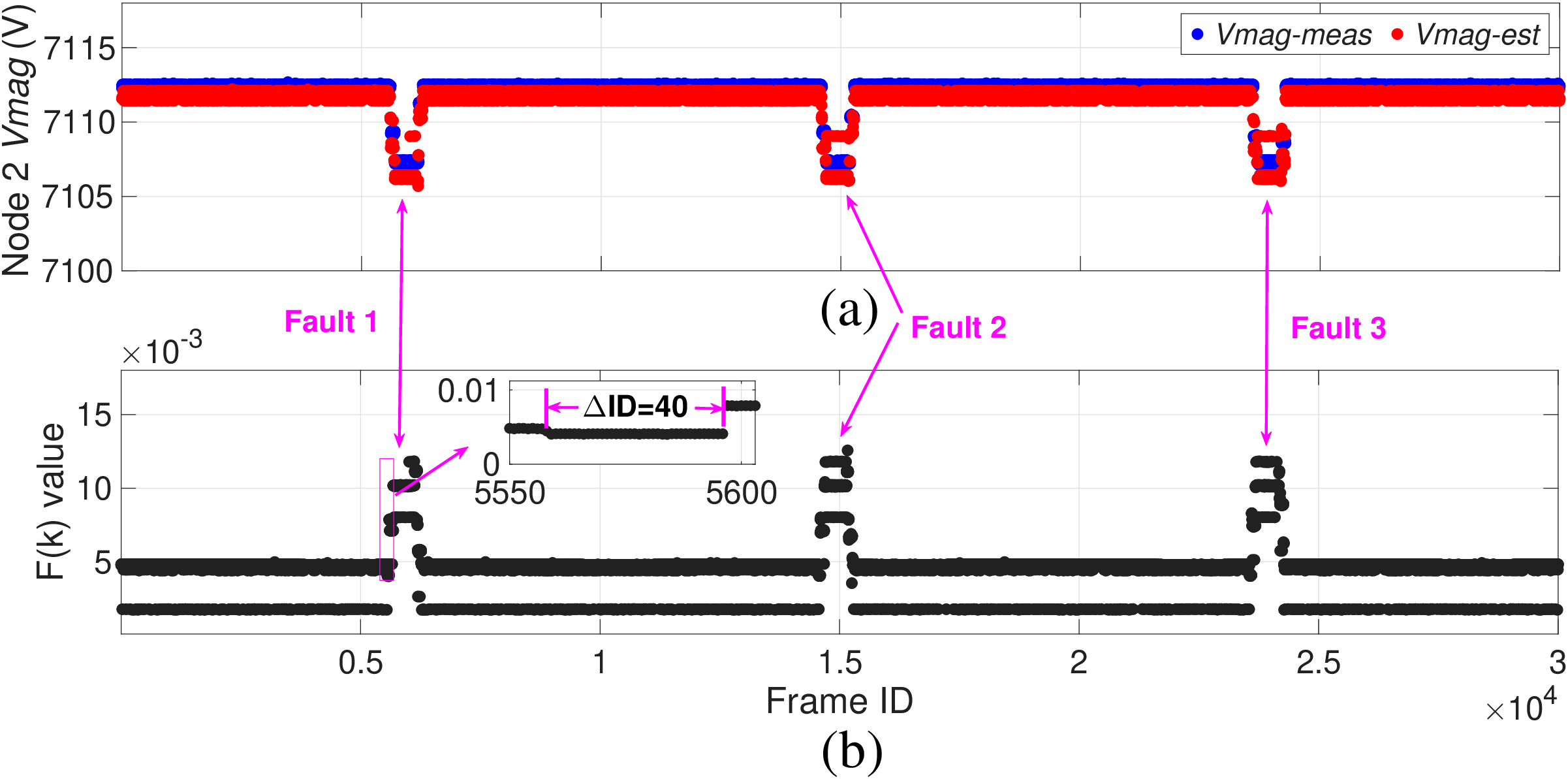}
    \caption{Fault-detection SE results: (a) measured vs.\ estimated $V_{mag}$ at Node 2 and (b) cost function values $F(k)$ over the frame ID.}
    \label{fig:OV_FE}
\end{figure}

\begin{table}
\centering
\caption{Fault Detection Delay Across the Faults.}
\label{tab:fault_detection_50hz}
\setlength{\tabcolsep}{1.8pt}
\begin{tabular}{ |c |c |c |c |c|}
\hline
Fault no. & {Fault insertion frame ID} & {Fault detection frame ID} & {$\Delta$ID} & {Delay (s)} \\
\hline
 1 & 5557 & 5597 & 40 & 0.80 \\
 2 & 14558 & 14598 & 40 & 0.80 \\
 3 & 23556 & 23600 & 44 & 0.88 \\

\hline
\end{tabular}
\label{tab:FE_accuracy}
\end{table}

Three fault events are injected at frame IDs 5557, 14558, and 23556 during the experiment. Fig.~\ref{fig:OV_FE}(a) shows the voltage magnitude at Node~2 with three injected faults highlighted in magenta. Once a fault is inserted, clear deviations appear between $Vmag\text{-}meas$ and $Vmag\text{-}est$ at Node~2, where the faults are applied.

Fig.~\ref{fig:OV_FE}(b) shows the cost function values $F(k)$ over the frame IDs with annotation of the injected faults. Each injected fault produces a sharp increase in $F(k)$ that clearly exceeds the threshold $T$. As illustrated in the zoomed inset of Fig.~\ref{fig:OV_FE}(b), Fault~1 is inserted at frame ID~5557 and detected at frame ID~5597, corresponding to a difference of 40 frames. Given a reporting rate of 50\,fps (0.02\,s per frame), this yields a detection delay of 0.80\,s. The detection delays for all three faults are summarized in Table~\ref{tab:FE_accuracy}.

It is important to note that the phasor data were measured and transmitted at 50\,fps, corresponding to a time resolution of 0.02\,s (20\,ms) per measurement. As discussed earlier, we used an external \acs{ADC} and, for the fault-detection experiments, employed all four channels---three to read the signals from each node and one to record the fault-trigger signal. Because reading multiple ADC channels introduces processing delays, and due to our system implementation, we could not operate reliably above 50\,fps while maintaining data transmission timing accuracy. As a result, detection delays are quantized in 20\,ms intervals, and any faster changes occurring between consecutive samples cannot be observed in the recorded data.

\section{Conclusion} \label{section:Conclusion_SE}

This paper presents a multi-node 5G-enabled \acs{SG} testbed integrated with a Typhoon \acs{HIL} real-time simulator, capable of transmitting real-time power system measurements over a commercial cellular network. The \acs{SG} nodes are implemented using \acs{RPi} devices equipped with 5G HAT modules to enable 5G connectivity. An extensive network performance evaluation was conducted on a U.S.-based commercial cellular network using KPIs such as delay, jitter, and frame loss to assess the suitability of 5G of \acs{SG} communications. 

The results demonstrate consistent performance across both indoor and outdoor environments and show that 5G can reliably accommodate a wide range of reporting rates, from as low as 0.5\,fps to as high as 120\,fps, corresponding to a broad spectrum of SG applications. The experimental results further indicate that 5G provides significantly lower delay, reduced jitter, and improved reliability compared to LTE cat-M. In particular, the highest mean delay observed for the 5G system was 32.66\,ms at a reporting rate of 0.5\,fps, which is approximately 6.5$\times$ lower than the corresponding LTE cat-M result at the same reporting rate.

We further experimentally validated real-time SE over a commercial 5G link using synchrophasor data from an IEEE 4-node feeder, evaluating estimation accuracy under both steady-state and dynamic load conditions. In addition, the proposed system demonstrates effective fault-detection capability, achieving detection delays as low as 0.80\,s. These results confirm the suitability of commercial 5G networks for real-world \acs{SG} monitoring, estimation, and protection tasks.  

The current testbed design relies on an external 16-bit \acs{ADC} to convert the Typhoon \acs{HIL} analog outputs into digital inputs for the \acs{RPi}-based \acs{SG} nodes. The accuracy of this conversion could be further improved by employing higher-precision ADC modules available on the market, which would likely enhance both load-tracking accuracy and fault-detection sensitivity.

\bibliographystyle{IEEEtran}
\bibliography{references}
\end{document}